# LINE DENSITY INDICE AS AN ALTERNATIVE TO MK PROCESS

M. Kuassivi
170374 Cotonou, Benin

ABSTRACT

73 broadband optical spectra of dwarf stars later than F0 have been obtained from the Nearby Stars Project website[1]. The number of absorption lines is computed for each spectrum between 6000 and 6200 Angstrom. A correlation is found between the density of lines $K_\lambda$ and the spectral type. This method is independent of calibration process, does not require high resolution or high signal—to—noise data and does not make use of a large library of standard spectra.

## 1. INTRODUCTION

Accurate spectral types provide basic physical parameters (effective temperature and surface gravity). They can also be used to pick out peculiar or astrophysically interesting stars.

In 1943, "An Atlas of Stellar Spectra" was produced at the Yerkes observatory. The Atlas represented a new approach to the classification of stars, one that depends on a set of standard "specimens". The Morgan—Keenan (MK) System of spectral classification is a classical application of morphological techniques (Keenan, 1984). Such techniques have a fundamental characteristics: stability over time (Garrison, 1994a).

Usually, the stars are classified on the MK system by direct visual inspection on a computer screen. To avoid this time consuming process, softwares are available that match a target spectrum to one from a standard dataset (see the ALLSTAR program described in Henry et al. 2002). This is not an easy task: calibrated fluxes are interpolated, spectra are normalized in a region free of opacity, spurious features must be rejected (telluric lines and cosmic rays). Although these routines can be tuned to work well with a particular dataset and a well—defined interval of spectral types, an universal procedure is yet to be written.

In Fig.1 are shown two normalized spectra in a limited wavelength region with different spectral types. The difference between the two spectra on the basis of the number of lines is obvious.

Such visual inspection cannot give an accurate spectral type for a given star: many lines can show up that are not seen in all spectra due to metallicity effect, spurious absorption lines may also appear in poor signal—to—noise data.

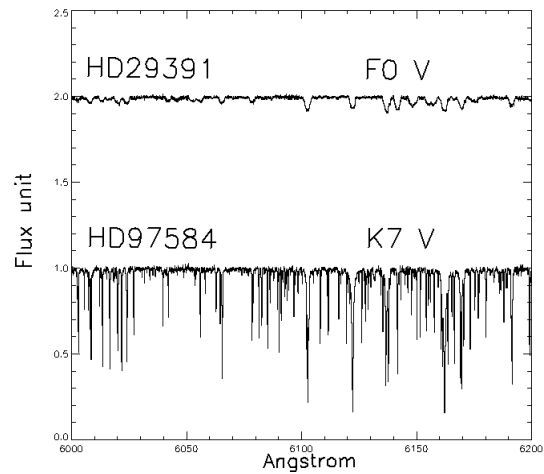

**FIG 1.** Comparizon of spectral type

However, the need for large computer routines comparing precisely flux calibrated spectra with a complete set of standard spectra over a large wavelength domain seems avoidable. Using a simple method, Gray et al. (2002) have shown that basic empirical indicators can be well suited to characterize giant stars.

---

[1] See http://bifrost.crwu.edu/nstars/

In this paper, I propose an alternative to a global morphological comparizon of standard spectra by focusing on one empirical parameter, independant of calibration : the density of lines detected in a given spectral range (hereafter $K_\lambda$).

| Spectral type | HD names | Spectral type | HD names |
|---|---|---|---|
| F0V | HD29391 | G4V | HD52711 |
| F2V | HD128167 | G4V | HD179958 |
| F3V | HD32715 | G5V | HD4208 |
| F5V | HD114378 | G5V | HD25680 |
| F5V | HD126141 | G5V | HD72946 |
| F5V | HD197692 | G5V | HD134987 |
| F6V | HD30652 | G5V | HD140538 |
| F6V | HD69897 | G5V | HD172051 |
| F6V | HD142860 | G6V | HD202206 |
| F7V | HD25998 | G7V | HD111395 |
| F7V | HD88595 | G8V | HD144579 |
| F7V | HD126660 | K0V | HD166 |
| F7V | HD143333 | K0V | HD10780 |
| F7V | HD215648 | K0V | HDD69830 |
| F8V | HD5015 | K0V | HD112758 |
| F8V | HD9826 | K0V | HD158633 |
| F8V | HD85380 | K0V | HD185144 |
| F8V | HD179949 | K1V | HD170657 |
| F9V | HD78366 | K1V | HD26965 |
| F9V | HD81858 | K1V | HD52698 |
| G0V | HD1461 | K2V | HD208801 |
| G0V | HD4614 | K2V | HD61606 |
| G0V | HD39587 | K2V | HD23356 |
| G0V | HD48682 | K2V | HD22049 |
| G0V | HD84117 | K2V | HD4628 |
| G0V | HD101177 | K2V | HD18445 |
| G0V | HD101563 | K3V | HD29697 |
| G0V | HD140913 | K3V | HD223778 |
| G0V | HD157214 | K3V | HD219134 |
| G0V | HD160269 | K3V | HD192310 |
| G0V | HD206860 | K3V | HD128165 |
| G1V | HD126053 | K3V | HD82106 |
| G1V | HD130948 | K3V | HD110833 |
| G1V | HD146233 | K3V | HD50281 |
| G2V | HD14802 | K4V | HD131977 |
| G3V | HD30495 | K4V | HD190007 |
| G4V | HD38529 | | |

**Table 1. List of targets as function of their spectral type**

## 2. OBSERVATIONS AND ANALYSIS

The observations reported in this paper are public data obtained through the Nearby Star Project web site (http://bifrost-.crwu.edu/nstars). The goal of the project is to seek information on the population of stars in the solar neighborhood. All the data have been observed with a single instrument at Mc Donald Observatory which makes the data set highly consistent (Heiter & Luck, 2003). Tables 1. lists the 73 stars used in this work as a function of their assigned spectral type.

The data analysis is straihgtforward. Spectra of many types were first carefully observed and the spectral domain between 6000 and 6200 Angstrom was chosen : smooth continuum and large variability of the number of lines with spectral types. Then I wrote a short I.D.L. routine to sort the number of lines in that domain that was robust enough to detect artifacts. The detection treshold was automatically adjusted for each spectrum depending on the signal—to—noise ratio (line depth between 80 and 92 per cent).

## 3. RESULTS AND DISCUSSION

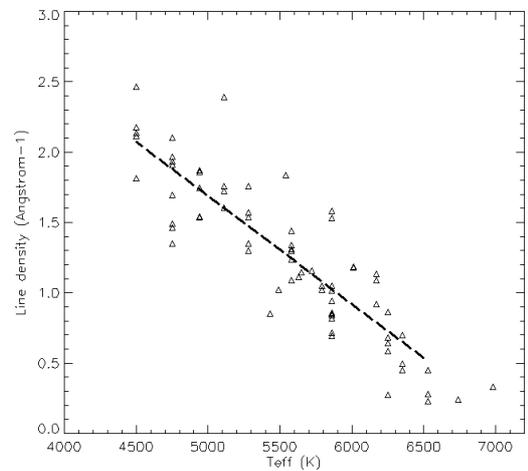

**Fig 2. Line density versus effective temperature. The dashed line is the linear regression.**

In Fig.2, I plot the line density indice $K_\lambda$ as a function of the effective temperature. A clear linear correlation is observed:

$$K_\lambda = -7.7\ 10^{-4}\ (0.5\ 10^{-4}) \times T_{eff} + 5.5\ (0.3)$$

Numbers in parenthesis are 1—sigma uncertainties on the parameters. Those uncertainties are relatively small in respect to the many uncertainties that can affect the data analysis: errors on the derived spectral type, lack of sophistication of the detection procedure, and intrinsic variability of the number of lines in various physical conditions.

Variation in the number of absorption lines is a trivial observation (Fig 1.) consistent with the increase of opacity with decreasing Effective Temperature. One that might have been neglected when assigning spectral type. This work shows that for late-type dwarf the number of lines gives a fair guess of the spectral type. The method can be improved in many ways: multi wavelength analysis, extension to earlier spectra types, to giant stars, more sophisticated routines for the detection of weak absorption lines and additional use of line ratios.

This work is an attempt to underline that a complete MK process reaches far beyond the scope of assigning spectral types (Garrison, 1994a; Garrison, 1994b; Walborn, 2008). The speed and capacity of current computers have led many to use big tools for small jobs. Short routines based on empirical parameters should be properly fitted for spectral classification.